%% file: Potts.tex
\documentclass[aps,prb,10pt,twocolumn,letterpaper,amsmath,amssymb]{revtex4-2}

\newcommand{\beginsupplement}{%
        \setcounter{section}{0}
        \setcounter{table}{0}
        \renewcommand{\thetable}{S\arabic{table}}%
        \setcounter{figure}{0}
        \renewcommand{\thefigure}{S\arabic{figure}}%
        \setcounter{equation}{0}
        \renewcommand{\theequation}{S\arabic{equation}}%
     } %\beginsupplement

\usepackage[utf8]{inputenc}
\usepackage{graphicx}
\usepackage{subfigure}
\usepackage[usenames,dvipsnames]{xcolor}
\usepackage{epstopdf}
\usepackage{amsmath,amsthm,amssymb}
\usepackage{tcolorbox}
\usepackage{latexsym}
\usepackage{bm}
\usepackage{dcolumn}
\usepackage{braket}
\usepackage[normalem]{ulem}
\usepackage{times}
\usepackage{hyperref}
\usepackage{enumitem}
\newcommand{\ignore}[1]{}
\newcommand{\red}[1]{\textcolor{black}{#1}}
\newcommand{\rred}[1]{\textcolor{black}{#1}}

\newcommand{\kb}{k_\mathrm{B}}
\newcommand{\tm}[1]{\mathbb{T}_{#1}}
\newcommand{\CNN}{\langle\delta(\sigma_i,\sigma_{i+1})\rangle}
\newcommand{\CNNN}{\langle\delta(\sigma_i,\sigma_{i+2})\rangle}

\begin{document}
%\begin{CJK*}{}{} % Use default fonts from CJK
%\bibliographystyle{apsrev4-1}
%\bibliographystyle{science}
%\bibliographystyle{naturemag}

%PRL Acceptance Criteria:
%1. Open a new research area, or a new avenue within an established area.
%2. Solve, or make essential steps towards solving, a critical problem.
%3. Introduce techniques or methods with significant impact.
%4. Be of unusual intrinsic interest to PRL's broad audience.

\title{Exact solution of the frustrated Potts model with next-nearest-neighbor interactions in one dimension via AI bootstrapping}
\author{Weiguo Yin}
\email{wyin@bnl.gov}
\affiliation{Condensed Matter Physics and Materials Science Division, Brookhaven National Laboratory, Upton, New York 11973, USA}

\begin{abstract}
The one-dimensional (1D) $J_1$-$J_2$ $q$-state Potts model is solved exactly for arbitrary $q$ 
\red{by analytically block-diagonalizing the original $q^2\times q^2$ transfer matrix into a simple $2\times 2$ maximally symmetric subspace,}
based on using OpenAI's reasoning model \texttt{o3-mini-high} to exactly solve the $q=3$ case. \red{Furthermore, by matching relevant subspaces, we map the Potts model onto a simpler effective 1D $q$-state Potts model, where $J_2$ acts as the nearest-neighbor interaction and $J_1$ as an effective magnetic field, nontrivially generalizing a 56-year-old theorem previously limited to the simplest case ($q=2$, the Ising model).}
Our exact results provide insights to phenomena such as atomic or electronic order stacking in layered materials and the emergence of dome-shaped phases \red{in complex phase diagrams}. %often seen in unconventional superconductors. 
This work is anticipated to fuel both research in 1D frustrated magnets for recently discovered finite-temperature application potentials and the fast moving topic area of AI  in science.
\bigskip
\\DOI: \href{https://doi.org/10.1103/y5vc-3t6q}{10.1103/y5vc-3t6q}
%\newpage
%\emph{One sentence summary:}
%A microscopic theory of entropic elasticity, structural stability, and negative thermal expansion (NTE) of a Coulomb floppy network describes unusual properties of a class of open framework crystals and shows remarkable agreement with experiment, thus uncovering the common origin of these properties in solid ceramics and soft matter, such as polymers and rubber.

\end{abstract}

{%PhySH:
%Interdisciplinary Physics => Chemical Physics \& Physical Chemistry => Thermodynamics
%Statistical Physics => Classical statistical mechanics
%Discipline(s) => Physical System(s) => Concept(s)
%Condensed Matter \& Materials Physics, Statistical Physics => Many-body techniques
}
\date{Received 31 March 2025; revised 6 June 2025; accepted 26 August 2025; published 11 September 2025}

\maketitle
%\end{CJK*}

%More than 12,000 papers have been published between 1969 and 1997 using the Ising model.

\section{Introduction}

Finding novel phases and phase transitions is a central challenge in various research fields, including condensed matter physics, materials science, quantum information, and microelectronics~\cite{Kivelson_24_book_statistical}. Unusual phases abound in frustrated magnets~\cite{Ramirez_review_frustrated_magnets_94,Balents_nature_frustration}, which are described typically by the Ising model %~\cite{Ising1925}
or the quantum Heisenberg model %~\cite{Heisenberg_1928}
with competing 
%spin-spin interactions either in the form of an equilateral triangle or via competition between the 
nearest-neighbor (NN) interaction $J_1$ and next-nearest-neighbor (NNN) interaction $J_2$~\cite{Kivelson_24_book_statistical,Ramirez_review_frustrated_magnets_94,Balents_nature_frustration,Yin_PRL_FeTe}. 

\rred{Another} fundamental model of statistical mechanics is the $q$-state Potts model~\cite{Potts_1952,Potts_RMP_82,Mattis_book_08_SMMS,Baxter_book_Ising}, which is a generalization of the Ising model ($q=2$) and can serve as a useful intermediary to study the transition from discrete (Ising) to continuous (Heisenberg) symmetry. 
%\red{For example, the Potts model with various $q$ values, say $q=21$, has been extensively used to study protein sequences~\cite{Potts_protein_PRE_13,Potts_PNAS_22_protein}.} %,Potts_protein_17,Potts_18_protein}. 
In particular, the one-dimensional (1D) $J_1$-$J_2$ Potts model could be relevant to problems ranging from the out-of-plane stacking of atomic or electronic orders in layered materials, such as charge stripe ordering in La$_{1.67}$Sr$_{0.33}$NiO$_{4}$~\cite{Yin_PRL_nicklate}, the Star-of-David charge-density wave in 1$T$-TaS$_2$~\cite{TaS2_PRL_19_DFT}, and spin spiral ordering in the Weyl semimetal EuAuSb~\cite{EuAuSb}\red{---where in-plane ordering is governed primarily by detailed quantum-mechanical effects, owing to in-plane interactions that are far stronger than those between planes---}to a time series with multiple choices at every time step such as table tennis training drill designs. 

While the $J_1$-$J_2$ Ising and Heisenberg models in one dimension~\cite{Dobson_JMathP_69_Many‐Neighbored-Ising-Chain,Stephenson_CanJP_70_J1-J2-Ising-chain,Fleszar_Baskaran_JPC_85_J1-J2,White_J1_J2} and two dimensions~\cite{Gangat_PRB_24_J1-J2_Ising_2D,Chandra_PRL_90_J1-J2_Heisenberg_Ising_transition,Roth_PRB_23_2D-J1-J2_ML-VMC} have been extensively studied, only the 1D $J_1$-$J_2$ Ising model has been solved exactly by using the transfer matrix method~\cite{Kramers_Wannier_PR_41_transferMatrix}. Exact solutions of the 1D $J_1$-$J_2$ Potts model remain unknown as well. The challenge arises from rapid increase in the order of the transfer matrix, which equals $q^2$. No wonder a $9\times 9$ matrix for $q=3$ is already hard to solve analytically and diagonalization of a $\left(10^{10^{10}}\right)^2\times \left(10^{10^{10}}\right)^2$ matrix for $q=10^{10^{10}}$ is simply beyond reach even numerically. \red{Previous studies remarkably reduced the task to numerical calculations for an effective $q\times q$ matrix in the integer-$q$ formalism of the transfer matrix---and for an effective $2\times 2$ matrix in the continuous-$q$ formalism of the transfer matrix where physics is less transparent---however, short of analytic exact results~\cite{Glumac_JPA_93_Potts_LR}. Hence, an intuitive understanding of the rich phase behaviors in the 1D $J_1$-$J_2$ Potts model is still lacking.} Since the model with $q=3$ already exhibits a distinct ground-state phase behavior from that with $q=2$ (cf. Fig.~\ref{Fig:PhaseDiagram}), it is of fundamental importance to exactly solve the model for arbitrary $q$. 

Two recent developments shed light on this long-standing problem. The first one is the analytic reduction of the $4\times 4$ transfer matrix for a decorated Ising ladder to an effective $2\times2$ matrix using symmetry-based block diagonalization, leading to the discovery of spontaneous finite-temperature ultranarrow phase crossover (UNPC), which exponentially approaches the forbidden finite-temperature phase transition in 1D Ising models~\cite{Yin_MPT}, and the subsequent discovery of in-field UNPC driven by exotic ice-fire states~\cite{Yin_MPT_chain,Yin_Ising_III_PRL,Yin_site_arXiv}. These findings point out the promising potentials of 1D frustrated magnets in finite-temperature applications; finding exact solutions for 1D frustrated Potts model could define a milestone in this important new direction. The second development is the derivation of an elegant equation, which determines the critical temperature of UNPC in decorated Ising models in an external magnetic field, by OpenAI's reasoning model \texttt{o3-mini-high} at the first-ever 1000-Scientist AI Jam Session~\cite{Yin_site_arXiv}. Hence, the author was inspired to prompt this AI reasoning model progressively and reflectively to handle the transfer matrix in \emph{the integer-$q$ formalism} for the $q=3$ case---despite quite a few errors in AI's responses---and eventually have found a symmetry-based block diagonalization that can analytically reduce the $9\times 9$ transfer matrix of the 1D $J_1$-$J_2$ three-state Potts model to an effective $2\times2$ matrix \red{(see Supplemental Material I~\cite{SI:J1_J2_Potts})}.

For general $q$, the key symmetry is the full permutation symmetry of the $q$ Potts states. In other words, the Hamiltonian (and therefore the transfer matrix in the integer-$q$ formalism) is invariant under any permutation of the labels $\{1,2,3,...,q\}$; its symmetry group is $\mathcal{S}_q$. Although the AI failed to go further but warned that the number of permutations increases dramatically as $q$ increases, the analytic results for the $q=2$ and $3$ cases---especially those that both arrive at an effective $2\times 2$ matrix---stimulated the author to realize that since only the largest eigenvalue ($\lambda$) of the transfer matrix matters in the thermodynamic limit, the task is reduced to identify the symmetry-separated subspace that contains $\lambda$. Then, the author found that this subspace is spanned by two maximally symmetric vectors, \red{according to the Perron–Frobenius theorem for positive matrices~\cite{Horn_Johnson_1985_Chapter8,Cuesta_1D_PT} (see Supplemental Material II)}. %(because all the transfer-matrix entries are exponentials of real energies---hence positive). 
The resulting $2\times 2$ matrix is easy to solve; therefore, the surprisingly simple exact solution of the 1D $J_1$-$J_2$ Potts model has been obtained for arbitrary $q$.

\red{The maximally symmetric subspace (MSS) method is in spirit similar to the effective Hamiltonian method for studying strongly correlated electronic materials~\cite{Zhang_Rice_Singlet,Yin_PRB_09_cuprates,Yin_PRL_Sr3CuIrO6}---with the effective model parameters depending on $J_1$, $J_2$, $q$, and temperature (see Supplemental Material III.A~\cite{SI:J1_J2_Potts}). Furthermore, a rare exact mapping between two canonical Hamiltonians %---with independent parameters---%
naturally emerges within the MSS method, namely the 1D $J_1$-$J_2$ $q$-state Potts model can be mapped onto a simpler 1D $q$-state Potts model, where $J_2$ acts as the NN interaction and $J_1$ as an effective magnetic field. %---thus reducing the order of the transfer matrix from $q^2$ to $q$. 
This not only nontrivially generalizes a 56-year-old theorem previously limited to $q=2$~\cite{Dobson_JMathP_69_Many‐Neighbored-Ising-Chain} but also profoundly bridges two research areas focused on spontaneous behaviors and external field effects, respectively, thereby demonstrating relevant-subspace-matching as a broadly applicable theoretical approach to tackling complex problems.}

\begin{figure}[b]
%\vspace{-0.5cm}
    \begin{center}
\includegraphics[width=0.9\columnwidth,clip=true,angle=0]{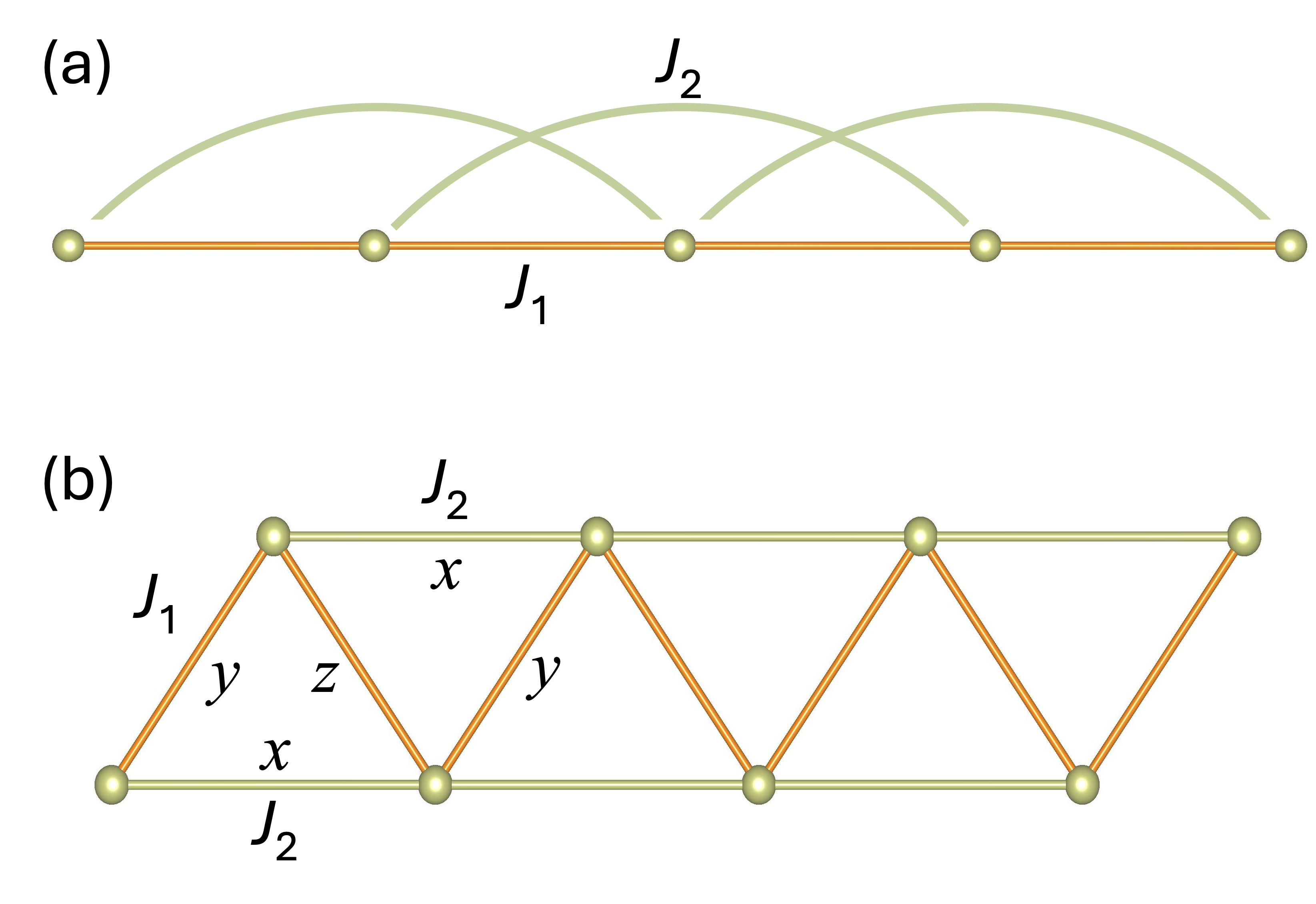}
    \end{center}
    \vspace{-0.5cm}
\caption{Schematics of (a) the single-chain $J_1-J_2$ Potts model~\cite{Potts_RMP_82} and (b) its equivalent zigzag-ladder representation. The balls stand for the spins with $q$ states. The orange bonds stand for the nearest-neighbor interactions $J_1$ and the greenish bonds stand for the next-nearest-neighbor interactions $J_2$. The contribution of each bond to the partition function is $x=e^{\beta J_2}$, $y=e^{\beta J_1/2}$, or $ z=e^{\beta J_1}=y^2$ when its two end spins are the same, and one otherwise.}
\label{Fig:structure}
\end{figure}

\section{Model and Solution}

We consider the following Hamiltonian [Fig.~1(a)]
\begin{eqnarray}
H=-J_1\sum_{i=1}^{2N}\delta({\sigma_{i},\sigma_{i+1}})-J_2\sum_{i=1}^{2N}\delta({\sigma_{i},\sigma_{i+2}}),
\label{minimal}
\end{eqnarray}
where $\sigma_i \in \{1,2,3,...,q\}$ is the spin variable at site $i$. $\delta({\sigma_{i},\sigma_{i+1}})$ is the Kronecker delta (which equals 1 if  $\sigma_{i}=\sigma_{i+1}$ and 0 otherwise). %$J_1$ and $J_2$ are the nearest-neighbor and next-nearest-neighbor interactions, respectively. 
$2N$ is the total number of the spins with $\sigma_{2N+1}\equiv\sigma_{1}$ and $\sigma_{2N+2}\equiv\sigma_{2}$, viz., the periodic boundary condition. 

To construct the transfer matrix, we use \red{both the overlapping-pairs formulation for Equation~(\ref{minimal}) with one spin per unit cell to get $\mathbb{T}$ and} the equivalent zigzag-ladder version of the model with two spins per unit cell [Fig.~1(b)] \red{to obtain $\mathbb{T}'$; they satisfy $\mathbb{T}'=\mathbb{T}^2$ (see Supplemental Material I~\cite{SI:J1_J2_Potts})}. In the thermodynamic limit $N\to\infty$, the partition function $Z=\mathrm{Tr}\,e^{-\beta H}=\lambda^{2N}$, where $\lambda$ is the largest eigenvalue of the  transfer matrix $\mathbb{T}$.  
The free energy per spin is given by
\begin{equation}
    f=\lim_{N\to\infty}-\frac{1}{2N\beta} \ln Z=-\frac{1}{\beta} \ln\lambda,
\end{equation} 
where $\beta=1/(k_\mathrm{B}T)$ with $T$ being the absolute temperature and $k_\mathrm{B}$ the Boltzmann constant. $f$ determines physical properties such as the entropy per spin $S=-\partial f/\partial T$, the specific heat $C_v=T\partial S/\partial T$, the NN correlation $\CNN=-\partial f/\partial J_1$, the NNN correlation $\CNNN=-\partial f/\partial J_2$, and the energy per spin $E=-\partial f/\partial \beta=-J_1\CNN-J_2\CNNN$. 

The $\lambda$-containing subspace is spanned by the following two maximally symmetric $q^2\times 1$ vectors:
\begin{eqnarray}
    |\psi_1\rangle &=& \frac{1}{\sqrt{q}}\sum_{s=1}^{q}\sum_{s'=1}^{q} \delta(s,s')|s,s'\rangle,\nonumber\\
    |\psi_2\rangle &=& \frac{1}{\sqrt{q^2-q}}\sum_{s=1}^{q}\sum_{s'=1}^{q}\left[1-\delta(s,s')\right]|s,s'\rangle.
    \label{eq:MSS}
\end{eqnarray}
The resulting transformation matrix $U_{q2}=\{|\psi_1\rangle, |\psi_2\rangle\}$ is a $q^2\times 2$ matrix that projects the $q^2\times q^2$ transfer matrix $\mathbb{T}$ to the $2\times2$ block $\mathbb{T}_{2}=U_{q2}^\top\,\mathbb{T}\,U_{q2}$, which is decoupled from the rest thanks to different symmetry and given by
\begin{equation}
    \mathbb{T}_{2}=\left(
\begin{array}{cc}
 u & w \\
 w & v \\
\end{array}
\right)=\left(
\begin{array}{cc}
 xy^2 & \sqrt{q-1}\,y \\
 \sqrt{q-1}\,y & x+q-2 \\
\end{array}
\right)
\label{eq:T2}
\end{equation}
%where 
%\begin{eqnarray}
%    u&=&x^2y^2z+(q-1)y^2, \nonumber\\
%   v&=&(x+q-2)^2+(q-1)z,\\
%    w&=&\sqrt{q-1} (x z+x+q-2)y, \nonumber
%    u&=&xy^2, \nonumber\\
%    v&=&x+q-2,\\
%    w&=&y\sqrt{q-1}, \nonumber
%\end{eqnarray}
with the shorthand notations
$x=e^{\beta J_2}$ and $y=e^{\beta J_1/2}$.
%, $ z=e^{\beta J_1}$. 
Note that the maximally symmetric subspace means that the expressions of $u$, $v$, and $w$ can be obtained straightforward by combinatorial analysis. The largest eigenvalue of $\mathbb{T}$ is the larger eigenvalue of $\mathbb{T}_{2}$ and is given by
\begin{equation}
    \lambda=\frac{u+v}{2}+\sqrt{\left(\frac{u-v}{2}\right)^2+w^2}.
    \label{eq:lambda}
\end{equation}

\section{Mapping}

\red{The 1D $J_1$-$J_2$ $q$-state Potts model can be mapped onto a simpler $q$-state Potts model, where $J_2$ acts as a NN interaction and $J_1$ appears as an effective magnetic field:
\begin{eqnarray}
H_\mathrm{eff}=-J_2\sum_{i=1}^{2N}\delta({\sigma_{i},\sigma_{i+1}})-J_1\sum_{i=1}^{2N}\delta({\sigma_{i},1}).
\label{eq:h}
\end{eqnarray}
Proof: By the Perron–Frobenius theorem~\cite{Horn_Johnson_1985_Chapter8}, the  largest eigenvalue of the $q\times q$ transfer matrix of Equation~(\ref{eq:h}) lives in the MSS spanned by the following two $q\times 1$ vectors:
\begin{eqnarray}
    |\phi_1\rangle &=& \sum_{s=1}^{q} \delta(s,1)|s\rangle,\nonumber\\
    |\phi_2\rangle &=& \frac{1}{\sqrt{q-1}}\sum_{s=1}^{q}\left[1-\delta(s,1)\right]|s\rangle.
    \label{eq:MSS}
\end{eqnarray}
The resulting $2\times 2$ matrix~\cite{Glumac_JPA_94_Potts_J1-h} is exactly the same as Equation~(\ref{eq:T2}). \qed}

\red{This mapping was previously demonstrated for the simplest case by exploiting a special
property unique to $q = 2$~\cite{Dobson_JMathP_69_Many‐Neighbored-Ising-Chain,Stephenson_CanJP_70_J1-J2-Ising-chain,Fleszar_Baskaran_JPC_85_J1-J2}. Here, its nontrivial generalization to arbitrary $q$ was achieved by matching the MSS of the two $q$-state Potts models. It implies that an antiferromagnetic Potts model in an external magnetic field---which is more tunable in practice---is as fundamental as a Potts model spontaneously frustrated by competing interactions (see Supplemental Material III.B~\cite{SI:J1_J2_Potts}).}

\rred{With hindsight, if the mapping for arbitrary $q$ had been conjectured, it could have been verified by numerical calculations. Thus, this conjecture would have provided a hint to the general analytical solution.}

\section{Results and Discussion}

\begin{figure}[t]
%\vspace{-0.5cm}
    \begin{center}
\includegraphics[width=\columnwidth,clip=true,angle=0]{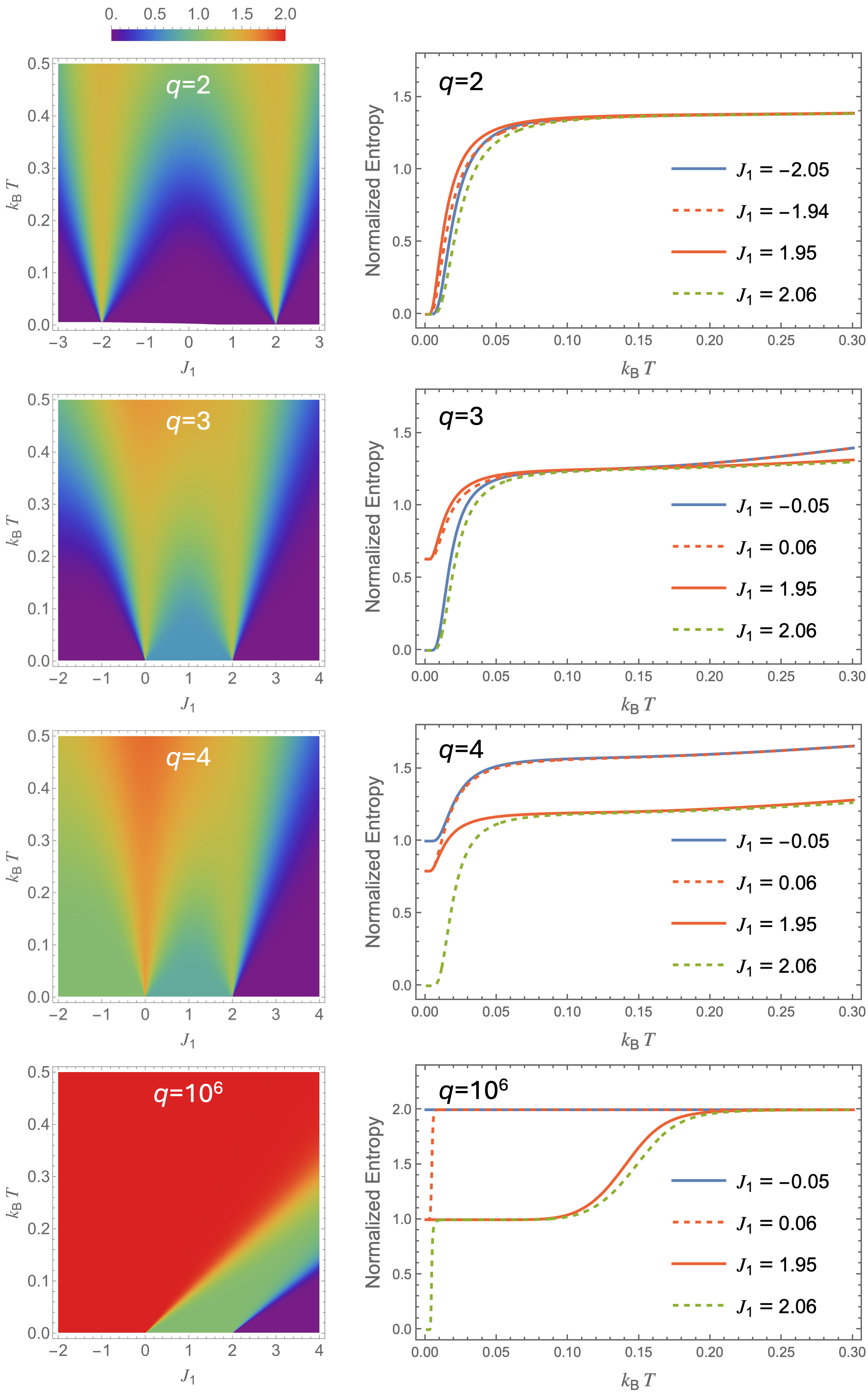}
    \end{center}
\vspace{-0.2cm}
\caption{Phase diagrams for $q=2, 3, 4, \mathrm{and}~10^6$. Left panels: Density plots of the normalized entropy $2S(J_1,T)/\ln q$ in the $J_1-T$ plane. Right panels: The temperature dependence of $2S(J_1,T)/\ln q$ for selected $J_1$ values around the critical points. $-J_2=1$ is set as the energy unit.}
\label{Fig:PhaseDiagram}
\end{figure}

The simplicity of Equation~(\ref{eq:T2}) provides an intuitive understanding of rich phase behaviors in the 1D $J_1$-$J_2$ Potts model. We are interested in frustrated cases where $J_2<0$ is antiferromagnetic and have set $-J_2=1$ as the energy unit from now on. The $q$-dependent phase diagrams given by the normalized entropy $2S(J_1,T)/\ln q$ are shown in Fig.~\ref{Fig:PhaseDiagram}. A $T_c$-dome-like phase emerges at low temperature for small $q$ and disappears for large $q$. The phase diagrams given by $\CNN$ and $\CNNN$ are shown in  Fig.~\ref{Fig:C12} in \rred{Appendix~B}.%End Matter.

To get insights to these rich phase diagrams, we start with analyzing ground-state phase behaviors. At $T=0$, the 1D $J_1$-$J_2$ Potts model for all $q$ values have three phases separated by two critical points (CPs), determined by the relative magnitude of $u$, $w$, $v$ in Equation~(\ref{eq:T2}). The right-hand-side phase for $J_1>2$ (where $u$ dominates) is always ``ferromagnetic'' with $\CNN=\CNNN=1$ and $E=-J_1-J_2$; more precisely, it is a mixture of $q$ ferromagnetic states, i.e., $\frac{1}{\sqrt{q}}(|111...\rangle+|222...\rangle+\cdots+|qqq...\rangle$. The ``ferromagnetic transition'' CP at $J_1=2$ has a residual entropy per spin 
\begin{equation}
    S(J_1=2,T=0)=\ln\left(\frac{1+\sqrt{4q-3}}{2}\right).
    \label{eq:entropyCP2}
\end{equation}
The $q=2$ case (i.e., the Ising model) differs from the $q\ge 3$ cases in two aspects: (i) the two CPs for $q=2$ are symmetry related and located at $J_1=\pm 2$, while they are located at $J_1=0$ and $2$ for $q\ge 3$. (ii) None of the three phases for $q=2$ has a macroscopic degeneracy, while one or two nontrivial states with residual entropy exist for $q\ge 3$. These phases and CPs will be elaborated in passing. Concerning the spin configurations in one dimension, we adopt the convention that the spins are arranged from left to right from now on.

$q=2$: The $J_1>0$ and $J_1<0$ regions are symmetry related by flipping the spins on one sublattice. Consequently, the left-hand-side phase for $J_1<-2$ is ``antiferromagnetic'' with $\CNN=0$, $\CNNN=1$, $E=-J_2$. The two CPs are located at $J_1=\pm 2$. The CP at $J_1=-2$ has the same residual entropy as that at $J_1=2$ given by Equation~(\ref{eq:entropyCP2}), i.e., $\ln(\frac{1+\sqrt{5}}{2})$, which is indeed the same as the magnetic-field-induced CP in the 1D Ising model with only NN interactions~\cite{Yin_g}. %, indicating the limited choice of the frustrated state for $q=2$.
The middle phase for $-2<J_1<2$ is a dimerized state, i.e., a pair of NN spins have the same variable but differ from the last pair, forming the patterns of ...11221122... and its three degenerate shifts, leading to $\CNN=1/2$, $\CNNN=0$, and $E=-J_1/2$. None of the three phases has a macroscopic degeneracy. 

$q\ge3$: The middle phase for $0<J_1<2$ (where $w$ dominates) is a randomly dimerized state (RDS) with $\CNN=1/2$, $\CNNN=0$, and $E=-J_1/2$, where like in the dimerized phase for $q=2$, a pair of NN spins have the same variable but differ from the last pair, so it has $q-1$ choices, yielding a residual entropy of $\ln(q-1)$ per two spins or $\ln\sqrt{q-1}$ per spin. This phase could be relevant to the stacking problem of the Star-of-David charge-density wave in 1$T$-TaS$_2$~\cite{TaS2_PRL_19_DFT}. The left-hand-side phase for $J_1<0$ (where $v$ dominates) is a paramagnet with $\CNN=\CNNN=0$ and $E=0$, where any spin differs from the two last spins, so it has $q-2$ choices, yielding a residual entropy of $\ln(q-2)$ per spin; notably, $q=3$ is special since the residual entropy is zero, forming the patterns of ...123123... and its five degenerate states by permutations of \{1, 2, 3\}, reminiscent of the ABC stacking pattern of the charge stripe order in La$_{1.67}$Sr$_{0.33}$NiO$_{4}$~\cite{Yin_PRL_nicklate} and the 120$^\circ$ spin spiral order in the Weyl semimetal EuAuSb~\cite{EuAuSb}. In the CP at $J_1=0$, any spin differs from its last next-nearest neighbor, so it has $q-1$ choices, yielding a residual entropy of $\ln(q-1)$ per spin; again, $q=3$ is special because its residual entropies at both CPs are equal to $\ln2$. 

\begin{figure}[t]
    \begin{center}
\includegraphics[width=0.95\columnwidth,clip=true,angle=0]{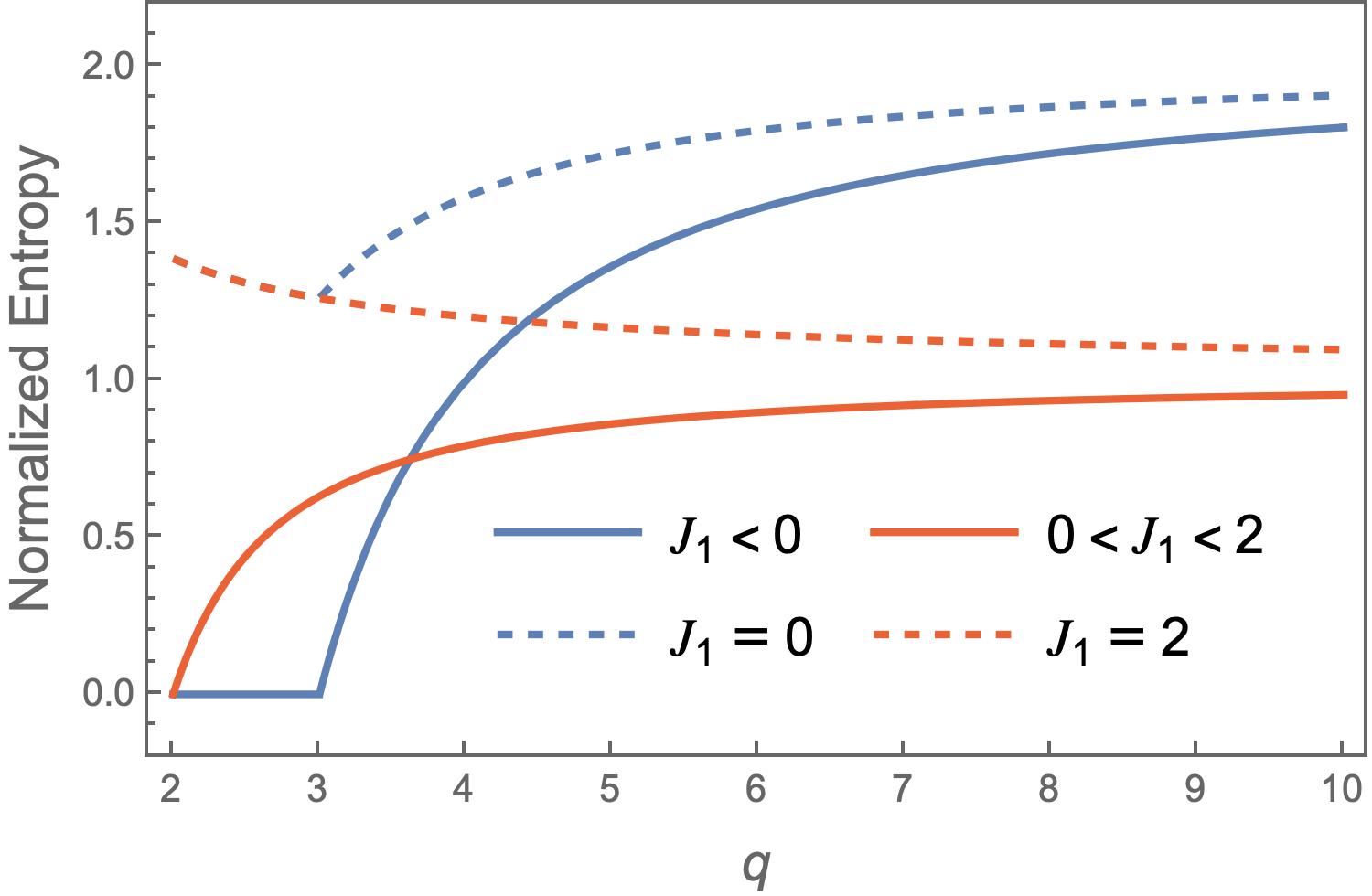}
    \end{center}
\caption{The $q\ge3$ dependence of the zero-temperature normalized entropy $2S(J_1,0)/\ln q$ for four different regions of $J_1$: $S=\ln(q-2)$ for $J_1<0$, $S=\ln(q-1)$ for $J_1=0$, $S=\ln\sqrt{q-1}$ for $0<J_1<2$, and $S=\ln[(1+\sqrt{4q-3})/2]$ for $J_1=2$. The $q=2$ case, which need be specially treated for different $J_1$ regions (see text), was added for comparison. $J_2=-1$.}
\label{Fig:QCP}
\end{figure}

The $q$ dependence of the residual entropies for the left and middle phases as well as the two CPs is summarized in Fig.~\ref{Fig:QCP}. For small $q$, the residual entropies of the CPs (dashed lines) are considerably larger than those of their adjacent phases (solid lines). Consequently, each CP develop a V-shape region in the $J_1-T$ phase diagram as $T$ increases (Fig.~\ref{Fig:PhaseDiagram}, left panels for $q=2,3,4$). The V-shape regions of the two CPs join to create a $T_c$-dome-like region for the middle randomly dimerized phase for $q\ge 3$. When placed near the CPs, the system does not follow the common lore---i.e., transition to the adjacent phase with higher macroscopic degeneracy---but transitions to the CP-developed V-shape region, also shown as the flat region in the $T$ curve of the entropy (Fig.~\ref{Fig:PhaseDiagram}, right panels for $q=2,3,4$), where the entropy value equals the corresponding CP's residual entropy.  

On the other hand, Fig.~\ref{Fig:QCP} reveals that for large $q$, the residual entropies of the CPs (dashed lines) approach those of one of their adjacent phases (solid lines), becoming indistinguishable---no more V-shaped CP regions (Fig.~\ref{Fig:PhaseDiagram}, left panel for $q=10^6$). When placed near a phase boundary, the system appears to follow the common lore, i.e., transition to the adjacent phase with higher macroscopic degeneracy. In particular, the low-temperature ferromagnetic phase for $J_1>2$ would undergo a two-step phase crossover, first to the middle randomly dimerized phase at 
%$\kb T\approx (J_1+2J_2)/\ln (q-1)$ 
$\kb T\approx (J_1+2J_2)/\ln q$ 
and then to the left paramagnetic phase at 
%$\kb T\approx J_1/[2\ln (q-2)-\ln(q-1)]$.
$\kb T\approx J_1/\ln q$.   

%The $T_c$ dome is a key phenomenon in phase diagrams of unconventional superconductivity in cuprates, iron-based superconductors, twisted bilayer graphene, etc.~\cite{Yin_PRB_phaseCompetition}. 

The $q$-dependent emerging and vanishing of a dome-like structure, controlled by the relative strength of the residual entropies of the phase's two CPs, provides one more mechanism for forming a dome-shaped phase, a key phenomenon in \red{complex materials, in addition to the two well-known} pictures of (i) a preformed order with phase coherence gradually building up~\cite{Emery_Nature_95} and (ii) competing phases~\cite{Chakravarty_Nature_04,Yin_PRB_phaseCompetition}. 

\section{Summary}

The 1D $J_1$-$J_2$ $q$-state Potts model has been solved exactly by discovering that the largest eigenvalue of the $q^2\times q^2$ transfer matrix lives in a $2\times 2$ maximally symmetric subspace\red{, which is equivalent to that of the simpler 1D $q$-state Potts model with $J_2$ acting as the NN interaction and $J_1$ as the magnetic field}. The model's ground state are found to feature three phases separated by two critical points for all $q$ values. The relative strength of the two critical points' residual entropies is large and small for small and large $q$, respectively, leading to the emerging and vanishing of a dome-shaped randomly dimerized phase for small and large $q$, respectively, providing a new mechanism for forming a dome-shaped phase and insights to the stacking problems of atomic or electronic orders in layered materials. These discoveries were based on using OpenAI's latest reasoning model \texttt{o3-mini-high} to exactly solve the $q=3$ case, which embodies a new paradigm of data- and value-driven discovery, inspiring scientists to evaluate and connect insights drawn from the vast---though imperfect---information provided by AI. 

\emph{Note added.} We recently employed the MSS method to demonstrate that the ''half ice, half fire'' driven ultranarrow phase crossover persists in 1D decorated Potts models for all $q>2$, exhibiting a $T_0$-dome structure in the %field-temperature 
phase diagram that is absent in the $q=2$ case~\cite{Yin_Potts_UNPC}.

\begin{figure*}[t]
%\vspace{-0.5cm}
    \begin{center}
\includegraphics[width=\textwidth,clip=true,angle=0]{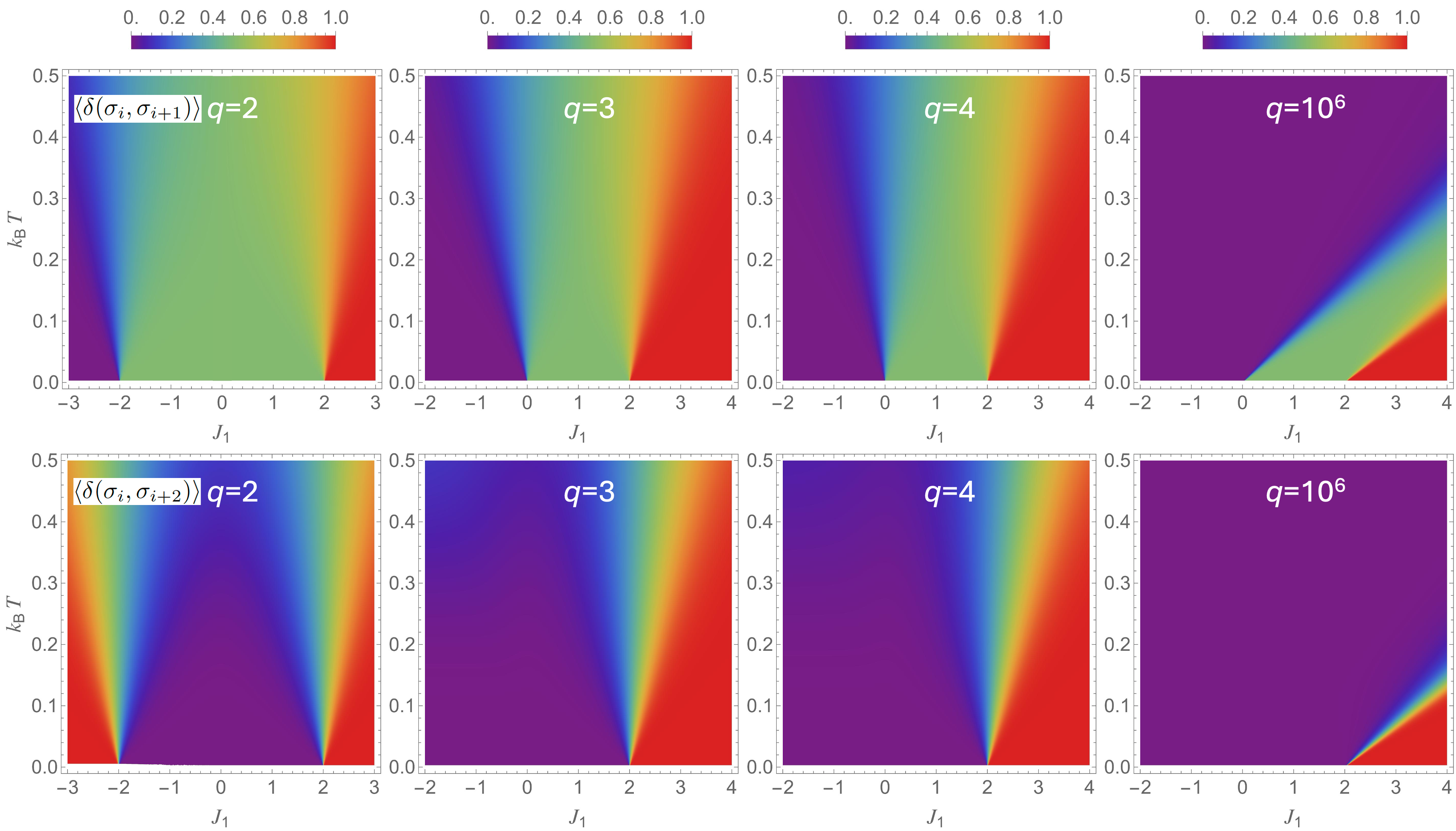}
    \end{center}
    \vspace{-0.3cm}
\caption{Phase diagrams in the $J_1-T$ plane for $q=2, 3, 4, \mathrm{and}~10^6$ given by the density plots of $\CNN$ (top panels) and $\CNNN$ (bottom panels). $-J_2=1$ is set as the energy unit.}
\label{Fig:C12}
\end{figure*}

\section*{Acknowledgment}

Brookhaven National Laboratory was supported by U.S. Department of Energy (DOE) Office of Basic Energy Sciences (BES) Division of Materials Sciences and Engineering under contract No. DE-SC0012704.

\section*{DATA AVAILABILITY}
The data that support the findings of this article are openly
available [37].

%\appendix

%\begin{widetext}
\section*{Appendix A: Software} 

\texttt{\uppercase{Wolfram Mathematica} 14.2} was used to plot Figs.~\ref{Fig:PhaseDiagram}--\ref{Fig:C12} and perform brute-force numerical computation of the eigenvalues of the $q^2\times q^2$ transfer matrix  for $q^2$ up to 1024 to verify the exact results. The codes together with that generated by the AI for the $q=3$ case are openly available~\cite{data:Wolfram1}. 
%in \url{https://community.wolfram.com/groups/-/m/t/3466026}. \;\;\; 

\texttt{VESTA 3.5.8}~\cite{VESTA} %(\url{https://jp-minerals.org/vesta/en/}) 
was used to plot Fig.~\ref{Fig:structure}. 

\vspace{0.5cm}
\section*{Appendix B: Additional phase diagrams\label{appendix:b}}

\red{Figure~\ref{Fig:C12} contains additional phase diagrams given by $\CNN$ and $\CNNN$. The zero-temperature phase transitions as a function of $J_1$ are first-order, since  $\CNN=-\partial f/\partial J_1$ is discontinuous at both $J_1=0$ and $J_1=2$; they as a function of negative $J_2$ the transitions are continuous at $J_1=0$ and first-order at $J_1=2$.}

%\end{widetext}

%\bibliography{Ising_quantum}
\input{Potts_v5.bbl}

\beginsupplement
\onecolumngrid
\newpage
\input{SI}

\end{document}

%% file: Potts_v5.bbl

%% file: SI.tex
%\begin{CJK*}{}{} % Use default fonts from CJK
%\bibliographystyle{apsrev4-1}
%\bibliographystyle{science}
%\bibliographystyle{naturemag}

%PRL Acceptance Criteria:
%1. Open a new research area, or a new avenue within an established area.
%2. Solve, or make essential steps towards solving, a critical problem.
%3. Introduce techniques or methods with significant impact.
%4. Be of unusual intrinsic interest to PRL's broad audience.

%\title{[Supplemental Material] Exact Solution of the Frustrated Potts Model with Next-Nearest-Neighbor Interactions in One Dimension: An AI-Bootstrapped Discovery}
\title{[Supplemental Material] Exact solution of the frustrated Potts model with next-nearest-neighbor interactions in one dimension via AI bootstrapping}
\author{Weiguo Yin}
\email{wyin@bnl.gov}
\affiliation{Condensed Matter Physics and Materials Science Division, Brookhaven National Laboratory, Upton, New York 11973, USA}

\date{Received 31 March 2025; revised 6 June 2025; accepted 26 August 2025; published 11 September 2025}

\maketitle
%\end{CJK*}

\beginsupplement
} %\ignore

\begin{center}
\textbf{\Large [Supplemental Material] Exact solution of the frustrated Potts model with next-nearest-neighbor interactions in one dimension via AI bootstrapping\\}

\textrm{\\Weiguo Yin\\}
%\email{wyin@bnl.gov}
\vspace{0.1cm}
\textsl{Condensed Matter Physics and Materials Science Division, Brookhaven National Laboratory, Upton, New York 11973, USA\\}
\textrm{(Received 31 March 2025; revised 6 June 2025; accepted 26 August 2025; published 11 September 2025)}
\end{center}

\section{AI's Exact Solution for $q=3$}

First of all, OpenAI's latest reasoning model \texttt{o3-mini-high} testified that to the best of its knowledge, the 1D $J_1$-$J_2$ Potts model had not been exactly solved. 

Next, the AI suggested that we used the overlapping-pairs formulation with \emph{one} spin per unit cell to construct the transfer matrix $\mathbb{T}$. However, anticipating that the AI would make mistakes, the author prompted the AI to use the equivalent zigzag-ladder version of the 1D $J_1$-$J_2$ three-state Potts model for two reasons: (i) The ladder version contains both NN and NNN interactions in one unit cell, so it would be easier to spot a mistake, if any, in constructing the transfer matrix $\mathbb{T}'$. (ii) The ladder version has \emph{two} spins per unit cell, hence
$\mathbb{T}'=\mathbb{T}^2$, which could be used to check the results. 

\subsection{The zigzag-ladder formulation}

The AI correctly gave the zigzag-ladder Hamiltonian
\begin{eqnarray}    
H_\mathrm{ladder} = -J_1 \sum_{n=1}^{N} \Bigl[
  \delta\bigl(\sigma_{n,A},\sigma_{n,B}\bigr)
  + \delta\bigl(\sigma_{n,B},\sigma_{n+1,A}\bigr)
\Bigr]
-J_2 \sum_{n=1}^{N} \Bigl[
  \delta\bigl(\sigma_{n,A},\sigma_{n+1,A}\bigr)
  + \delta\bigl(\sigma_{n,B},\sigma_{n+1,B}\bigr)
\Bigr],
\label{eq:ladder}
\end{eqnarray}
with periodic boundary conditions 
\(\sigma_{N+1,A}\equiv\sigma_{1,A}\) and \(\sigma_{N+1,B}\equiv\sigma_{1,B}\). Here, $A$ and $B$ denote the two legs. $\sigma_{n,A} \in \{1,2,3\}$ is the spin variable at site $n$ on the $A$ leg.

\ignore{
\begin{figure}[b]
%\vspace{-0.5cm}
    \begin{center}
\includegraphics[width=0.5\columnwidth,clip=true,angle=0]{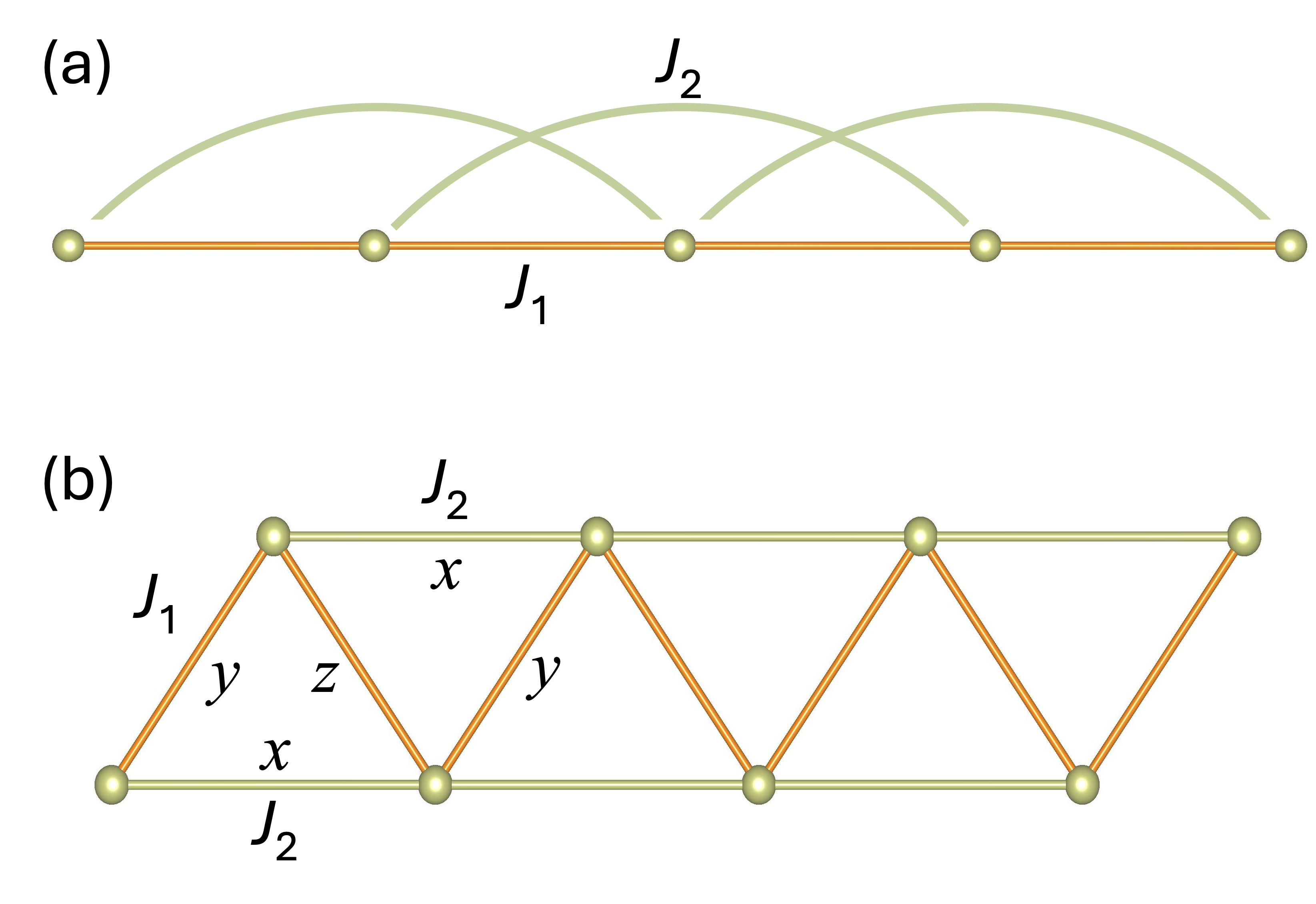}
    \end{center}
    \vspace{-0.5cm}
\caption{Schematics of (a) the single-chain $J_1-J_2$ Potts model~\cite{Potts_RMP_82} and (b) its equivalent zigzag-ladder representation. The balls stand for the spins with $q$ states. The orange bonds stand for the nearest-neighbor interactions $J_1$ and the greenish bonds stand for the next-nearest-neighbor interactions $J_2$. The contribution of each bond to the partition function is $x=e^{\beta J_2}$, $y=e^{\beta J_1/2}$, or $ z=e^{\beta J_1}=y^2$ when its two end spins are the same, and one otherwise.}
\label{Fig:structure}
\end{figure}
}

The AI also correctly generated the following expression of the $q^2\times q^2$ transfer matrix
\begin{eqnarray}
        \mathbb{T}'\Bigl((a,b),(a',b')\Bigr) &=& \exp\!\Biggl\{ \frac{\beta J_1}{2}\Bigl[\delta(a,b) + \delta(a',b')\Bigr]
+\beta J_1\,\delta(b,a')
+\beta J_2\Bigl[\delta(a,a') + \delta(b,b')\Bigr] \Biggr\} \nonumber\\
&=& x^{\delta(a,a') + \delta(b,b')}\, y^{\delta(a,b) + \delta(a',b')}\,z^{\delta(b,a')}
\label{transfer}
\end{eqnarray}
where $(a,b)$ with $a,b\in\{1,2,3\}$ is a ``rung'' state consisting of a spin pair on a rung, and $(a',b')$ is a neighboring rung state. Let the 9 states of a rung state be ordered as follows: $(1, 1), (1, 2), (1, 3), (2, 1), (2, 2), (2, 3), (3, 1), (3, 2), (3, 3)$. With the shorthand notations
$x=e^{\beta J_2}$, $y=e^{\beta J_1/2}$, $ z=e^{\beta J_1}=y^2$, the transfer matrix is given explicitly by
$$
\mathbb{T}'=\left(
\begin{array}{ccccccccc}
 x^2 y^2 z & x y z & x y z & x y & y^2 & y & x y & y & y^2 \\
 x y & x^2 & x & z & x y z & z & 1 & x & y \\
 x y & x & x^2 & 1 & y & x & z & z & x y z \\
 x y z & z & z & x^2 & x y & x & x & 1 & y \\
 y^2 & x y & y & x y z & x^2 y^2 z & x y z & y & x y & y^2 \\
 y & 1 & x & x & x y & x^2 & z & z & x y z \\
 x y z & z & z & x & y & 1 & x^2 & x & x y \\
 y & x & 1 & z & x y z & z & x & x^2 & x y \\
 y^2 & y & x y & y & y^2 & x y & x y z & x y z & x^2 y^2 z \\
\end{array}
\right)
$$
However, the AI made mistakes in generating the above matrix form of $\mathbb{T}'$. It was prompted to make sure $\tm{27}'=\tm{34}'=\tm{48}'=\tm{62}'=\tm{76}'=\tm{83}'=1$. Then, the AI corrected its mistakes and identified the $\mathcal{S}_3$ symmetry group.

Then, the AI was prompted to block-diagonalize $\mathbb{T}'$. It found that $\mathbb{T}'$ can be block diagonalized to $\tm{\mathrm{block}}'=U^\top\,\mathbb{T}'\,U$ with the transformation matrix 
$$
U=\left(
\begin{array}{ccccccccc}
 \frac{1}{\sqrt{3}} & 0 & -\frac{1}{\sqrt{2}} & 0 & 0 & 0 & -\frac{1}{\sqrt{6}} & 0 & 0 \\
 0 & \frac{1}{\sqrt{6}} & 0 & 0 & -\frac{1}{\sqrt{2}} & -\frac{1}{\sqrt{6}} & 0 & 0 & -\frac{1}{\sqrt{6}} \\
 0 & \frac{1}{\sqrt{6}} & 0 & -\frac{1}{\sqrt{2}} & 0 & 0 & 0 & -\frac{1}{\sqrt{6}} & \frac{1}{\sqrt{6}} \\
 0 & \frac{1}{\sqrt{6}} & 0 & 0 & 0 & 0 & 0 & \sqrt{\frac{2}{3}} & \frac{1}{\sqrt{6}} \\
 \frac{1}{\sqrt{3}} & 0 & 0 & 0 & 0 & 0 & \sqrt{\frac{2}{3}} & 0 & 0 \\
 0 & \frac{1}{\sqrt{6}} & 0 & 0 & 0 & \sqrt{\frac{2}{3}} & 0 & 0 & -\frac{1}{\sqrt{6}} \\
 0 & \frac{1}{\sqrt{6}} & 0 & 0 & \frac{1}{\sqrt{2}} & -\frac{1}{\sqrt{6}} & 0 & 0 & -\frac{1}{\sqrt{6}} \\
 0 & \frac{1}{\sqrt{6}} & 0 & \frac{1}{\sqrt{2}} & 0 & 0 & 0 & -\frac{1}{\sqrt{6}} & \frac{1}{\sqrt{6}} \\
 \frac{1}{\sqrt{3}} & 0 & \frac{1}{\sqrt{2}} & 0 & 0 & 0 & -\frac{1}{\sqrt{6}} & 0 & 0 \\
\end{array}
\right)
$$
\ignore{
$$
\mathbb{T}_\mathrm{block}'=\left(
\begin{array}{ccccccccc}
 y^2 \left(x^2 z+2\right) & \sqrt{2} y (x z+x+1) & 0 & 0 & 0 & 0 & 0 & 0 & 0 \\
 \sqrt{2} y (x z+x+1) & x^2+2 x+2 z+1 & 0 & 0 & 0 & 0 & 0 & 0 & 0 \\
 0 & 0 & y^2 \left(x^2 z-1\right) & \frac{1}{2} y (x (2 z-1)-1) & \frac{1}{2} y (x (2 z-1)-1) & \frac{1}{2} \sqrt{3} (x-1) y & 0 & -\frac{1}{2} \sqrt{3} (x-1) y & 0 \\
 0 & 0 & -\frac{1}{2} y (x (z-2)+1) & x^2-\frac{z}{2}-\frac{1}{2} & \frac{x-z}{2} & \frac{1}{2} \sqrt{3} (z-x) & \frac{1}{2} \sqrt{3} y (x z-1) & \frac{1}{2} \sqrt{3} (z-1) & 0 \\
 0 & 0 & -\frac{1}{2} y (x (z-2)+1) & \frac{x-z}{2} & x^2-\frac{z}{2}-\frac{1}{2} & -\frac{1}{2} \sqrt{3} (z-1) & -\frac{1}{2} \sqrt{3} y (x z-1) & \frac{1}{2} \sqrt{3} (x-z) & 0 \\
 0 & 0 & \frac{1}{2} \sqrt{3} y (x z-1) & \frac{1}{2} \sqrt{3} (z-x) & \frac{1}{2} \sqrt{3} (z-1) & x^2-\frac{z}{2}-\frac{1}{2} & -\frac{1}{2} y (x (z-2)+1) & \frac{x-z}{2} & 0 \\
 0 & 0 & 0 & \frac{1}{2} \sqrt{3} (x-1) y & -\frac{1}{2} \sqrt{3} (x-1) y & \frac{1}{2} y (x (2 z-1)-1) & y^2 \left(x^2 z-1\right) & \frac{1}{2} y (x (2 z-1)-1) & 0 \\
 0 & 0 & -\frac{1}{2} \sqrt{3} y (x z-1) & -\frac{1}{2} \sqrt{3} (z-1) & \frac{1}{2} \sqrt{3} (x-z) & \frac{x-z}{2} & -\frac{1}{2} y (x (z-2)+1) & x^2-\frac{z}{2}-\frac{1}{2} & 0 \\
 0 & 0 & 0 & 0 & 0 & 0 & 0 & 0 & (x-1)^2 \\
\end{array}
\right)
$$
}
While $\mathbb{T}_\mathrm{block}'$ is too lengthy to display here, the first $2\times 2$ block of the resulting block-diagonalized transfer matrix is given by
$$
\Big(\mathbb{T}_\mathrm{block}'\Big)_{2\times 2}=\left(
\begin{array}{cc}
 y^2 \left(x^2 z+2\right) & \sqrt{2} y (x z+x+1) \\
 \sqrt{2} y (x z+x+1) & (x+1)^2+2 z \\
\end{array}
\right)=
\left(
\begin{array}{cc}
 xy^2 & \sqrt{2}\, y \\
 \sqrt{2}\, y & x+1 \\
\end{array}
\right)^2\; \mathrm{for}\; z=y^2,
$$
whose larger eigenvalue $\lambda'$ is the largest eigenvalue of the transfer matrix $\mathbb{T}'$. Indeed, $(\mathbb{T}_\mathrm{block}')^{}_{2\times 2}$ is a square of another matrix.

After that, the AI was prompted to generate the \texttt{Wolfram Mathematica 14.2} code for the above conversation. 
%The task was done in seconds with little need of correction. 
However, the AI failed to generate a workable \texttt{Mathematica} code for $q>3$. Instead, it warned that the number of permutations in the $\mathcal{S}_q$ symmetry group increases dramatically as $q$ increases. 
%When pushed, the AI created a few fake \texttt{Mathematica} functions! %which ``might be worth implementing,'' said Yi Yin, Associate Director of Academic Innovation of Wolfram Research, Inc., at its exhibition booth in APS Global Summit 2025.

Later on, the similarity between $(\mathbb{T}_\mathrm{block}')^{}_{2\times 2}$ for $q=2$ and that for $q=3$ stimulated the author to realize that $(\mathbb{T}_\mathrm{block}')^{}_{2\times 2}$ is the maximally symmetric subspace of $\mathbb{T}'$. Brute-force numerical computations of the eigenvalues of the $q^2 \times q^2$ transfer matrix for $q^2$ up to 1024 were performed to  verify the exact results.

\ignore{\begin{equation}
    \lambda = 
    \frac{1}{2} \left(x^2+2 x+2 z+1+y^2 \left(x^2 z+2\right)\right)+\frac{1}{2} \sqrt{\left(x^2+2 x+2 z+1-y^2 \left(x^2 z+2\right)\right)^2+4 \left(\sqrt{2} y (x z+x+1)\right)^2}
\label{eq:lambda}
\end{equation}
}

\ignore{
% Requires: \usepackage{booktabs}
\begin{table}[b]
    \centering
 \caption{The elements of the $2\times 2$ block of the block diagonalized transfer matrix.}
    
    \begin{tabular}{c|c|c|c}
        \hline\hline
        $q$ & (11) & (12)=(21) & (22) \\
        \hline
        2   & $y^2 \left(x^2 z+1\right)$ & $ y (x z+x)$   & $x^2+z$   \\
        3   & $y^2 \left(x^2 z+2\right)$ & $\sqrt{2} y (x z+x+1)$ & $(x+1)^2+2 z$ \\
        4   & $y^2 \left(x^2 z+3\right)$ & $\sqrt{3} y (x z+x+2)$  & $(x+2)^2+3 z$ \\
        10  & $y^2 \left(x^2 z+9\right)$ & $3 y (x z+x+8)$  & $(x+8)^2+9 z$ \\
        q  & $y^2 \left(x^2 z+q-1\right)$ & $\sqrt{q-1} y (x z+x+q-2)$  & $(x+q-2)^2+(q-1)z$ \\
        \hline\hline
    \end{tabular}

    \label{table:block}
\end{table}
}

\subsection{The overlapping-pairs formulation}

Finally, we double checked the $q=3$ result using the overlapping-pairs formulation of the transfer matrix, which is given by
\begin{eqnarray}
    \mathbb{T}\Bigl((a,b),(a',b')\Bigr) &=& \delta(b,a') \exp\!\Biggl\{ \frac{\beta J_1}{2}\Bigl[\delta(a,b) + \delta(a',b')\Bigr]
+\beta J_2\delta(a,b') \Biggr\} \nonumber\\
   &=& \delta(b,a')\, x^{\delta(a,b')}\,y^{\delta(a,b)}\,\left(\sqrt{z}\right)^{\delta(a',b')}.
\label{transfer}   
\end{eqnarray}
The AI missed $\delta(b,a')$ and the resulting incorrect $\mathbb{T}$ did not satisfy $\mathbb{T}'=\mathbb{T}^2$. The AI corrected the error after being told that something was wrong. Then, it correctly generated the explicit matrix form:
$$
\mathbb{T}=\left(
\begin{array}{ccccccccc}
 x y \sqrt{z} & y & y & 0 & 0 & 0 & 0 & 0 & 0 \\
 0 & 0 & 0 & x & \sqrt{z} & 1 & 0 & 0 & 0 \\
 0 & 0 & 0 & 0 & 0 & 0 & x & 1 & \sqrt{z} \\
 \sqrt{z} & x & 1 & 0 & 0 & 0 & 0 & 0 & 0 \\
 0 & 0 & 0 & y & x y \sqrt{z} & y & 0 & 0 & 0 \\
 0 & 0 & 0 & 0 & 0 & 0 & 1 & x & \sqrt{z} \\
 \sqrt{z} & 1 & x & 0 & 0 & 0 & 0 & 0 & 0 \\
 0 & 0 & 0 & 1 & \sqrt{z} & x & 0 & 0 & 0 \\
 0 & 0 & 0 & 0 & 0 & 0 & y & y & x y \sqrt{z} \\
\end{array}
\right)
$$
In the same maximally symmetric subspace,
$$
\Big(\mathbb{T}_\mathrm{block}\Big)_{2\times 2}=
\left(
\begin{array}{cc}
 xy\sqrt{z} & \sqrt{2}\, y \\
 \sqrt{2}\, \sqrt{z} & x+1 \\
\end{array}
\right)=\left(
\begin{array}{cc}
 xy^2 & \sqrt{2}\, y \\
 \sqrt{2}\, y & x+1 \\
\end{array}
\right)\; \mathrm{for}\; z=y^2,
$$
Indeed, $\mathbb{T}'=\mathbb{T}^2$ and $(\mathbb{T}_\mathrm{block}')^{}_{2\times 2}=[(\mathbb{T}_\mathrm{block})^{}_{2\times 2}]^2$ for $z=y^2$. 
For simplicity, this formulation is used in the main text.

\section{Proof of $\lambda$ living in the Maximally Symmetric Subspace}

The Perron–Frobenius theorem for positive matrices states~\cite{Horn_Johnson_1985_Chapter8}: 

If $A$ is an $n\times n$ real matrix with all entries $A_{ij}>0$, then
\begin{enumerate}
    \item A has a real eigenvalue $\lambda > 0$ (the Perron root) which is strictly larger in modulus than all other eigenvalues.
    \item The eigenspace corresponding to $\lambda$ is one‐dimensional (1D), and there is a unique (up to scale) eigenvector $\mathbf{v}$ with strictly positive components.
\end{enumerate}
Because $\mathbf{v}$ is the only eigenvector with all-positive entries, it is forced to be fixed by any symmetry of $A$. Equivalently, if $G$ is any group of permutation matrices (or orthogonal transformations) commuting with $A$, then $g \mathbf{v}=\mathbf{v}$ for all $g\in G$, so $\mathbf{v}$ lies in the trivial representation of $G$ and is therefore the most symmetric of the eigenvectors.
In short, the Perron–Frobenius theorem guarantees a unique largest eigenvalue with a strictly positive eigenvector. Positivity of that eigenvector forces it to be invariant under every permutation symmetry of the matrix. Hence the Perron vector is the most symmetric among all eigenvectors.

This implies that the Perron vector is a superposition of the maximally symmetric vectors that can be conveniently obtained from the following procedure:
\begin{enumerate}
\item Starting with any basis state of $A$, add with equal weight the states obtained from all permutation-symmetry operations over the initial state, and assign zero weight to all the other states, to construct an $n\times 1$ vector $\mathbf{v}_1$.
\item Pick any basis state of $A$ that has zero weight in $\mathbf{v}_1$ and repeat the first step to construct an $n\times 1$ vector $\mathbf{v}_2$.
\item Pick any basis state of $A$ that has zero weight in all $\mathbf{v}_1$, $\mathbf{v}_2$, $\dots$, $\mathbf{v}_{m-1}$, repeat the previous step to construct an $n\times 1$ vector $\mathbf{v}_m$, and so on.
\item Stop when each basis state of $A$ has appeared with nonzero weight in one of $\mathbf{v}_1$, $\mathbf{v}_2$, $\dots$, $\mathbf{v}_{m}$.
\end{enumerate}
These orthonormal vectors form a $m$-dimensional maximally symmetric subspace (MSS). Define the transformation matrix $U=\{\mathbf{v}_1, \mathbf{v}_2, \dots, \mathbf{v}_{m}\}$; then, the $m\times m$ matrix $U^{\top}A \,U$ guarantees to contain the Perron vector and $\lambda$ as the largest eigenvalue in the MSS.

Here for the 1D 
$J_1$-$J_2$ Potts model, since all the transfer-matrix entries are exponentials of real energies---hence positive, the Perron-Frobenius is applicable with $n=q^2$ and $m=2$. We demonstrated that instead of directly solving the $q^2\times q^2$ transfer matrix for the Perron vector, it is much easier to first construct a 2D block-diagonalized subspace spanned by the two maximally symmetric vectors that can be conveniently identified. Then, the Perron vector is a superposition of the two maximally symmetric vectors. This MSS guarantees to contain $\lambda$. \qed

%It is straightforward to extend the MSS method in the integer-$q$ formalism of the transfer matrix to 1D Potts models with longer-range interactions:  $J_i$ the $i$th nearest-neighbor interaction and $J_L$ the longest. Then, the dimension of the MSS increases with $L$; its upper bound is the order of the transfer matrix in the continuous-$q$ formalism for the 1D $J_1$-$J_2$-$\cdots$-$J_{L-1}$ Potts model (with the interaction range being shorter by 1)~\cite{Glumac_JPA_93_Potts_LR}---it is $q$-independent for $q\ge L$. 

Note that the Perron–Frobenius theorem was employed to prove the nonexistence of any phase transition in 1D Ising models with short-range interactions~\cite{Cuesta_1D_PT}. 
%\end{widetext}

\section{The MSS method for Effective Hamiltonians}

Identifying a simpler effective Hamiltonian $H_\mathrm{eff}$ for the starting Hamiltonian $H$ is a widely pursued theoretical strategy for tackling complex problems~\cite{Zhang_Rice_Singlet,Yin_PRB_09_cuprates,Yin_PRL_Sr3CuIrO6}. This is typically done by deriving $H_\mathrm{eff}$ approximately from $H$. The resulting model parameters of $H_\mathrm{eff}$ are usually not independent but ``renormalized'' as they depend on the model parameters of $H$. When $H_\mathrm{eff}$ becomes significant enough, it may gain its own right to be considered as a starting or canonical Hamiltonian (i.e., its model parameters are independent and do not depend on $H$), thus opening a new research area. For example, the derivation of the $t$-$J$ model from a multi-band Hubbard model is valid in a limited parameter space of the Hubbard model: $J\propto t^2/U$ for the Hubbard interaction $U \gg t$. Yet, the $t$-$J$ model gains its own right as a canonical Hamiltonian in research for high-temperature superconductivity~\cite{Zhang_Rice_Singlet,Yin_PRB_09_cuprates}; thus, the rigorous mapping between the Hubbard model and the $t$-$J$ model is lost.

\subsection{Mapping onto an effective Ising model}

The present MSS method for the 1D $J_1$-$J_2$ $q$-state Potts model is in spirit similar to the reduction of $H$ to $H_\mathrm{eff}$, except that the MSS method is exact. Indeed, the $2\times2$ transfer matrix in the MSS, $\left(
\begin{array}{cc}
 u & w \\
 w & v \\
\end{array}
\right)
$, can be expressed as an effective 1D Ising model in an external magnetic field via the partition function $Z=e^{A}\,\mathrm{Tr}\,e^{-\beta H_\mathrm{eff}}$, where $A=\frac{1}{2}\left[\tfrac12(\ln u + \ln v)+\ln w\right]$, and the effective Hamiltonian is given by~\cite{Yin_MPT_chain,Yin_Ising_III_PRL}
\begin{eqnarray}
H_\mathrm{eff}=-J_\mathrm{eff}\sum_{i=1}^{2N}\sigma_{i}\sigma_{i+1}-h_\mathrm{eff}\sum_{i=1}^{2N}\sigma_{i}, 
\label{eq:Ising_eff}
\end{eqnarray}
where $\sigma_i=\pm1$ is the Ising spin values, and $h_\mathrm{eff}=\frac{1}{2\beta}\left(\ln u -\ln v\right)$ and $J_\mathrm{eff}=\frac{1}{2\beta}\left[\tfrac12(\ln u + \ln v)-\ln w\right]$. Apparently, $J_\mathrm{eff}$ and $h_\mathrm{eff}$ depend on $J_1$, $J_2$, $q$, and $T$. 

\subsection{Mapping between two canonical Hamiltonians}

Furthermore, we found that the 1D $J_1$-$J_2$ $q$-state Potts model can be mapped onto a simpler 1D $q$-state Potts model, where $J_2$ acts as a nearest-neighbor interaction and $J_1$ appears as an effective magnetic field:
\begin{eqnarray}
H_\mathrm{q}=-J_2\sum_{i=1}^{2N}\delta({\sigma_{i},\sigma_{i+1}})-J_1\sum_{i=1}^{2N}\delta({\sigma_{i},1}). 
\label{eq:h}
\end{eqnarray}
Its $q\times q$ transfer matrix is given by
\begin{eqnarray}
    \mathbb{T}_q\Bigl(a,b\Bigr) &=& \exp\!\Biggl\{\beta J_2\delta(a,b)+ \frac{\beta J_1}{2}\Bigl[\delta(a,1) + \delta(b,1)\Bigr]\Biggr\} = x^{\delta(a,b)}\,y^{\delta(a,1)+\delta(b,1)}.
\label{transfer}   \nonumber
\end{eqnarray}
In the explicit matrix form,
$$
\mathbb{T}_q=\left(
\begin{array}{ccccccccc}
  xy^2 & y & y & y & \cdots & y \\
     y & x & 1 & 1 & \cdots & 1 \\
     y & 1 & x & 1 & \cdots & 1 \\
     y & 1 & 1 & x & \cdots & 1 \\
\vdots & \vdots & \vdots &  \vdots & \vdots  & \vdots \\
     y & 1 & 1 & 1 & \cdots & x \\
 \end{array} \nonumber
\right).
$$
Its MSS is spanned by the following two $q\times 1$ vectors:
\begin{eqnarray}
    |\phi_1\rangle &=& \sum_{s=1}^{q} \delta(s,1)|s\rangle,\nonumber\\
    |\phi_2\rangle &=& \frac{1}{\sqrt{q-1}}\sum_{s=1}^{q}\left[1-\delta(s,1)\right]|s\rangle.
    \label{eq:MSS} \nonumber
\end{eqnarray}
The resulting $2\times 2$ matrix in the MSS is 
\begin{equation}
    \mathbb{T}_{2}=\left(
\begin{array}{cc}
 xy^2 & \sqrt{q-1}\,y \\
 \sqrt{q-1}\,y & x+q-2 \\
\end{array}
\right), \nonumber
\label{eq:T2}
\end{equation}
which is exactly the same as the $2 \times 2$ transfer matrix in the MSS for the 1D $J_1$-$J_2$ $q$-state Potss model [see Equation~(4) in the main text].

Dobson was the first to demonstrate this mapping for the simplest case ($q=2$, the Ising model) by exploiting a special property unique to $q=2$~\cite{Dobson_JMathP_69_Many‐Neighbored-Ising-Chain}. For $q=2$, the 1D $J_1$-$J_2$ Potts model can be recast to the standard $J_1$-$J_2$ Ising model:
\begin{eqnarray}
H=-\frac{J_1}{2}\sum_{i=1}^{2N}\sigma_{i}\sigma_{i+1}-\frac{J_2}{2}\sum_{i=1}^{2N}\sigma_{i}\sigma_{i+2}, %\nonumber
\label{eq:Ising}
\end{eqnarray}
where $\sigma_i=\pm1$ is the Ising spin values.
With the substitution of $\tau_i=\sigma_i\sigma_{i+1}$ and $\tau_i\tau_{i+1}=\sigma_i\sigma_{i+2}$, one obtains
\begin{eqnarray}
H_{q=2}=-\frac{J_2}{2}\sum_{i=1}^{2N}\tau_{i}\tau_{i+1}-\frac{J_1}{2}\sum_{i=1}^{2N}\tau_{i}, %\nonumber
\label{eq:Isingh}
\end{eqnarray}
where $\tau_i=\pm1$ is the Ising spin values~\cite{Dobson_JMathP_69_Many‐Neighbored-Ising-Chain,Stephenson_CanJP_70_J1-J2-Ising-chain,Fleszar_Baskaran_JPC_85_J1-J2}.
However, the above substitution does not apply to $q>2$. 

Here, the nontrivial generalization of the mapping to arbitrary $q$ was done by matching the MSS of the two $q$-state Potts models. 
%Here, a nontrivial generalization of Dobson's theorem to arbitrary $q$---thus reducing the order of the transfer matrix from $q^2$ to $q$---naturally emerges within the present MSS method.
This is a rare exact mapping between two canonical Hamiltonians for two research areas focused on spontaneous behaviors and external field effects, respectively.  
It implies that an antiferromagnetic Potts model in an external magnetic field---which is more tunable in practice---is as fundamental as a Potts model spontaneously frustrated by competing interactions.

%\bibliography{Ising_quantum}
%\input{Potts_v2.bbl}
%\end{document}